\documentclass[a4paper,12pt,twoside,fleqn]{article}
\usepackage{epsfig,times,citesort}
\usepackage[rflt]{floatflt}

\usepackage{tweaklist} % Better control of parameters in
\normalsize % Set \baselineskip
\newlength{\narrowfigwidth}
\makeatletter
\renewcommand{\section}{\@startsection{section}%
  {1}{0pt}{\baselineskip}{\baselineskip}{\bfseries\uppercase}}
\renewcommand{\subsection}{\@startsection{subsection}%
  {2}{0pt}{\baselineskip}{\baselineskip}{\bfseries\em}}
\makeatother

\newcommand{\captionfonts}{\small}

\makeatletter  % Allow the use of @ in command names
\long\def\@makecaption#1#2{%
  \vskip\abovecaptionskip
  \sbox\@tempboxa{{\captionfonts #1: #2}}%
  \ifdim \wd\@tempboxa >\hsize
    {\captionfonts #1: #2\par}
  \else
    \hbox to\hsize{\hfil\box\@tempboxa\hfil}%
  \fi
  \vskip\belowcaptionskip}
\makeatother   % Cancel the effect of \makeatletter

\begin{document}

\begin{center}\bf
  ATOMIC-SCALE MODELING OF THE DEFORMATION OF NANOCRYSTALLINE METALS
\end{center}
J. SCHI{\O}TZ$^{1,2,\ast}$, T. VEGGE$^{1,2}$, K. W. JACOBSEN$^1$\\
$^1$Center for Atomic-scale Materials Physics and Department of
Physics, Technical University of Denmark, DK-2800 Lyngby, Denmark\\
$^2$Materials Research Department, Ris{\o} National Laboratory,
DK-4000 Roskilde, Denmark.\\
$^\ast$Email: schiotz@fysik.dtu.dk

\section{abstract}

Nanocrystalline metals, i.e.\ metals with grain sizes from 5 to 50 nm,
display technologically interesting properties, such as dramatically
increased hardness, increasing with decreasing grain size.  Due to the
small grain size, direct atomic-scale simulations of plastic
deformation of these materials are possible, as such a polycrystalline
system can be modeled with the computational resources available
today.

We present molecular dynamics simulations of nanocrystalline copper
with grain sizes up to 13 nm.  Two different deformation mechanisms
are active, one is deformation through the motion of dislocations, the
other is sliding in the grain boundaries.  At the grain sizes studied
here the latter dominates, leading to a softening as the grain size is
reduced.  This implies that there is an ``optimal'' grain size, where
the hardness is maximal.

Since the grain boundaries participate actively in the deformation, it
is interesting to study the effects of introducing impurity atoms in
the grain boundaries.  We study how silver atoms in the grain
boundaries influence the mechanical properties of nanocrystalline
copper.

\section{Introduction}

In recent years, advanced production techniques have made it possible
to create metals, alloys and ceramics with grain sizes down to 5 nm.
For metals, this represents a reduction of the grain size by
approximately four orders of magnitude compared to most conventionally
produced metals.  As can be expected, such a dramatic change in the
microstructure leads to significant changes in the mechanical
properties of the metals.  For example, the hardness of typical
nanocrystalline metals is far higher than what is seen in their
coarse-grained counterparts \cite[and references
therein]{SiFo94,MoMo97}.

Nanocrystalline metals are an attractive group of metals to model, as
the small grain size provides a ``cut-off'' of the typical length
scales, where structures appear during deformation.  In coarse-grained
materials structures appear on vastly different length scales, making
it very difficult to model the properties of these materials.  The
models must include processes that occur on length scales from the
sub-nanometer scale of the atomic processes in dislocation cores, to
the micro- or even millimeter scale of grain and subgrain structures
\cite{CaLeLoPeSoWi98,CaTh98}.  Atomic-scale simulations of systems of
these sizes are beyond the reach of even the most powerful of todays
supercomputers.  One is thus forced to split the system into
sub-problems at different length-scales, and only treat sub-problems
at the atomic scale with atomistic models.  On coarser length scales
other modeling paradigms must be used, such as Dislocation Dynamics
and continuum plasticity calculations.

Dividing the problem into sub-problems at different length scales
often results in a better understanding of the problem, as it draws
attention to the structures which are relevant at a given length scale
(for example, when studying the formation of dislocation structures it is
clearly more relevant to focus on dislocations as the fundamental
concept rather than on individual atoms).  On the other hand, one
becomes dependent on this understanding, when creating the
coarse-grained models, as many assumptions about the relevant
phenomena at different length-scales will by necessity be built into
the multi-scale model.  The low grain size of nanocrystalline metals
``compresses'' this range of length scales to a range, where the whole
deformation problem can be modeled at the atomic scale, as many grains
in the polycrystalline material can be handled in an atomic-scale
simulation.  This makes it possible to perform unbiased
simulations of the deformation process, where no \emph{a priori}
assumptions are made about the deformation mechanisms.

In recent papers, we have presented simulations of the plastic
deformation of nanocrystalline Cu and Pd
\cite{ScDiJa98,ScVeDiJa98,ScVeDiJa98bx}.  Other authors have presented
simulations of the structure and elastic properties of nanocrystalline
metals and semiconductors
\cite{Ch95,PhWoGl95,PhWoGl95b,ZhAv96,KePhWoGl97}, and of the plastic
deformation of Ni under constant stress loading \cite{SwCa97,SwCa97b}.
In this paper we review our simulations of the deformation mechanisms
in nanocrystalline metals, and present simulations of the effects of
impurities in the grain boundaries.

\section{Simulations of pure metals}

\subsection{Simulation setup}

\begin{floatingfigure}{\narrowfigwidth}
  \begin{center}
    \epsfig{file=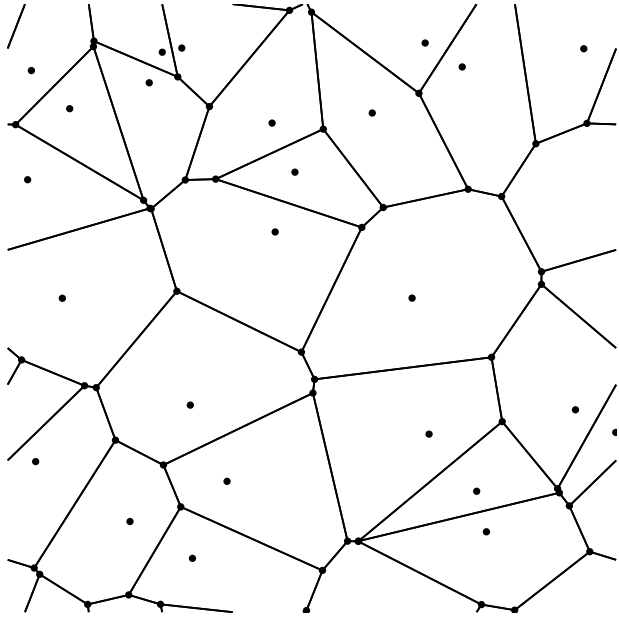, width=\narrowfigwidth}
  \end{center}
  \caption{An example of a two-dimensional ``grain structure''
    generated by a Voronoi construction.  Periodic boundary conditions
    are used.  A two-dimensional nanocrystalline material could be
    obtained by filling each cell with atoms in a regular lattice.
    A three-dimensional version of this algorithm is used when generating
    the systems studied in this paper.}
  \label{fig:voronoi}
\end{floatingfigure}
We have attempted to generate three-dimen\-sional systems with a
microstructure similar to that observed experimentally in samples
generated by inert gas condensation (IGC).  The grains appear to be
randomly oriented, approximately equiaxed, and dislocation free.  The
grain size distribution is close to log-normal \cite[and references
therein]{SiFo94,Si94b}.  We try to create systems that match this
description.

The grains are produced using a three-dimensional Voronoi tesselation:
random grain centers are chosen, and space is divided into regions in
such a way that each region consists of the points in space closer to
a given grain center than to any other grain center.  Each region is
then filled with a randomly oriented fcc lattice.  An example of a
two-dimensional Voronoi tessellation is shown in
Figure \ref{fig:voronoi}.

The generated samples are annealed for 50\,ps at 300\,K to relax the
grain boundary structure.  We found that the annealing time and
temperature are uncritical, but that the properties of the system are
different if \emph{no} annealing is done.

The interactions between the atoms are modeled using the Effective
Medium Theory (EMT) \cite{JaNoPu87,JaStNo96}.  EMT is a many-body
potential providing a realistic description of the metallic bonding,
in particular in  face-centered cubic (fcc) metals and alloys of fcc
metals.  Computationally, EMT is not much more demanding than pair
potentials, but provide a significantly more realistic description of
the metallic bonding.

The systems were deformed using a molecular dynamics (MD) procedure.
A conventional MD simulation was performed, but at each timestep the
atomic coordinates in the pulling direction (the $z$ coordinates) were
rescaled in order to deform the system gradually.  At the same time,
the box dimensions in the transverse directions are allowed to shrink
to keep the transversal components of the stress ($\sigma_{xx}$ and
$\sigma_{yy}$) close to zero \cite{ScVeDiJa98bx}.  The change in
system dimensions is slow at the timescale of the simulation, the
relative elongation is $2.5 \times 10^{-6}$ each timestep.  With a
timestep of 5 fs, this nevertheless results in a very high strain rate
($\dot\varepsilon = 5 \times 10^8 s^{-1}$).  The results presented
here are not very sensitive to the strain rate, although some
dependence is seen (typical stresses increase by 20\% when the strain
rate is changed from $2.5 \times 10^{7} s^{-1}$ to $1 \times 10^{9}
s^{-1}$) \cite{ScVeDiJa98bx}.

\subsection{Results}

\begin{figure}[tp]
  \begin{center}
    \leavevmode
    \epsfig{file=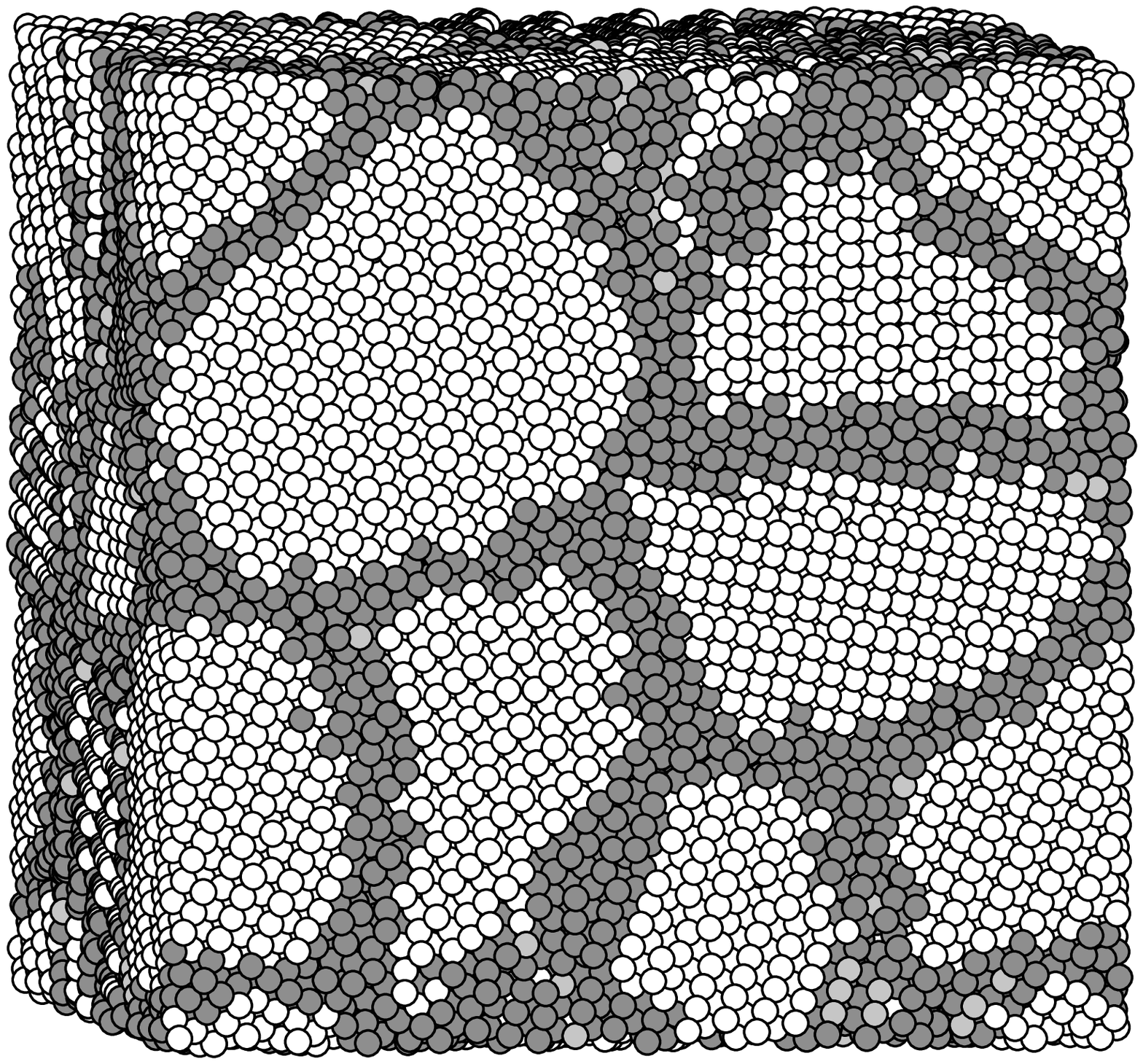, bb=42 56 511 534, clip=,
      width=0.49\linewidth}\hfill
    \epsfig{file=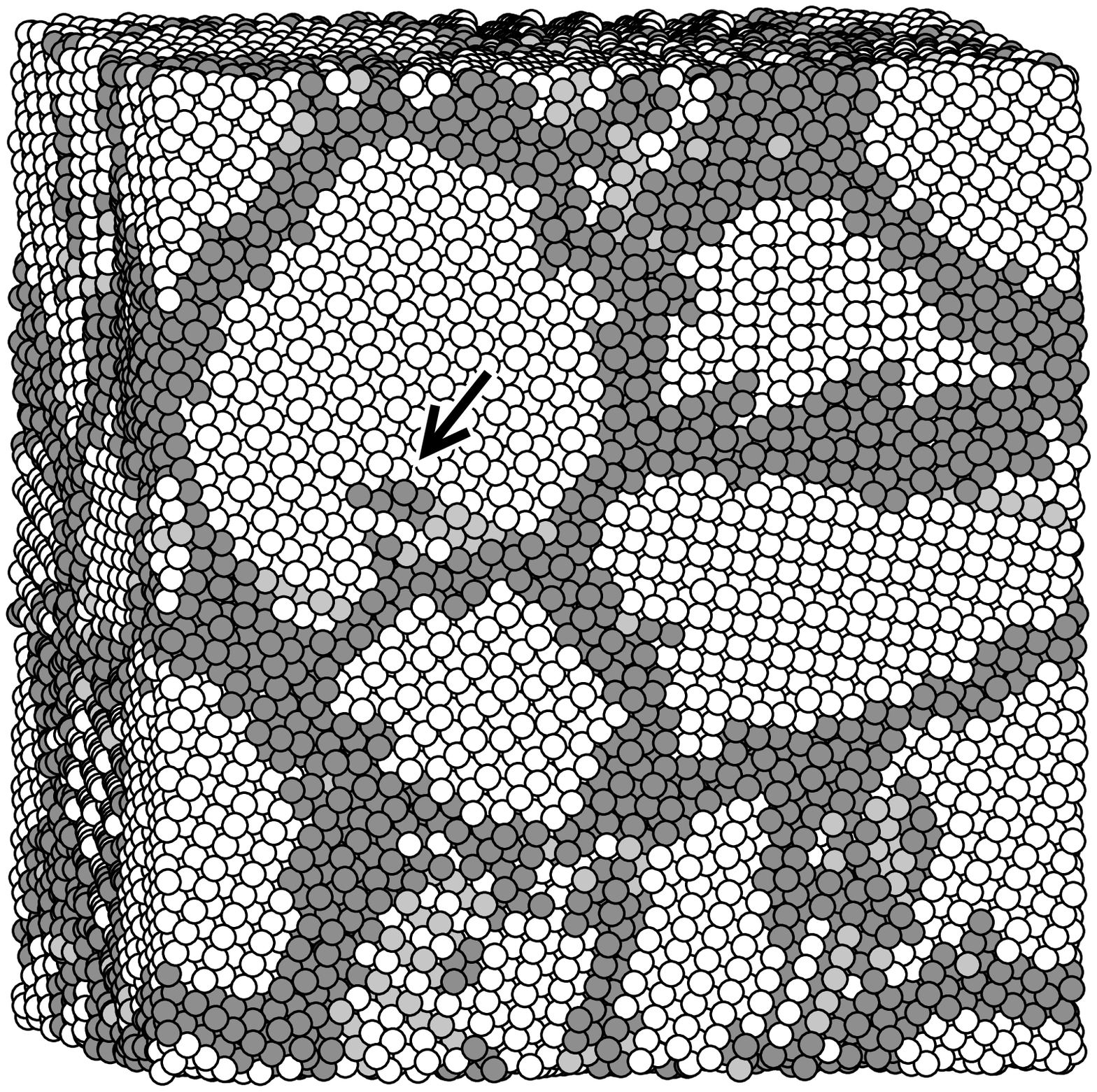, bb=42 56 511 534, clip=,
      width=0.49\linewidth}\\[2mm]
    \epsfig{file=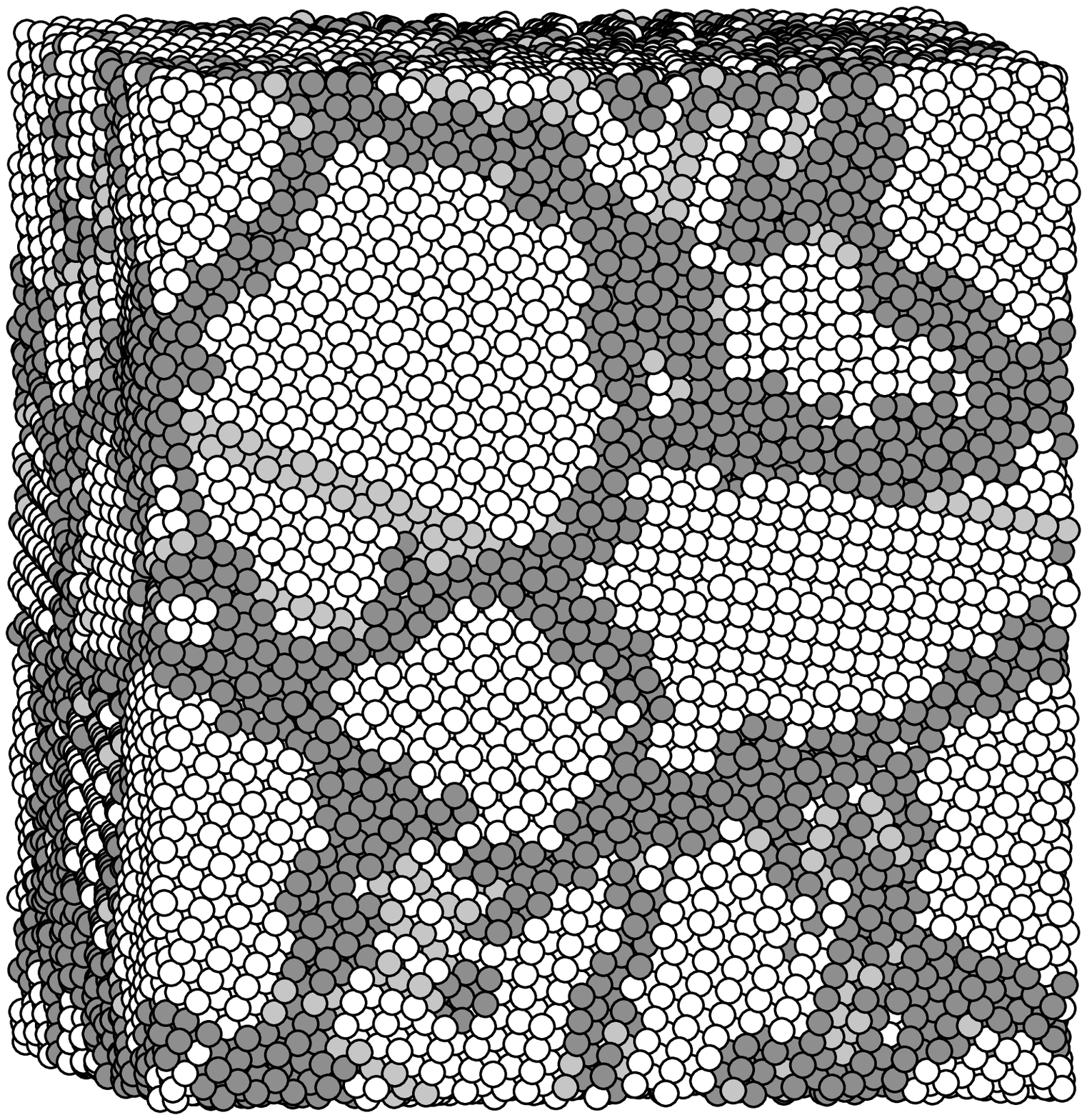, bb=42 56 511 534, clip=,
      width=0.49\linewidth}\hfill
    \epsfig{file=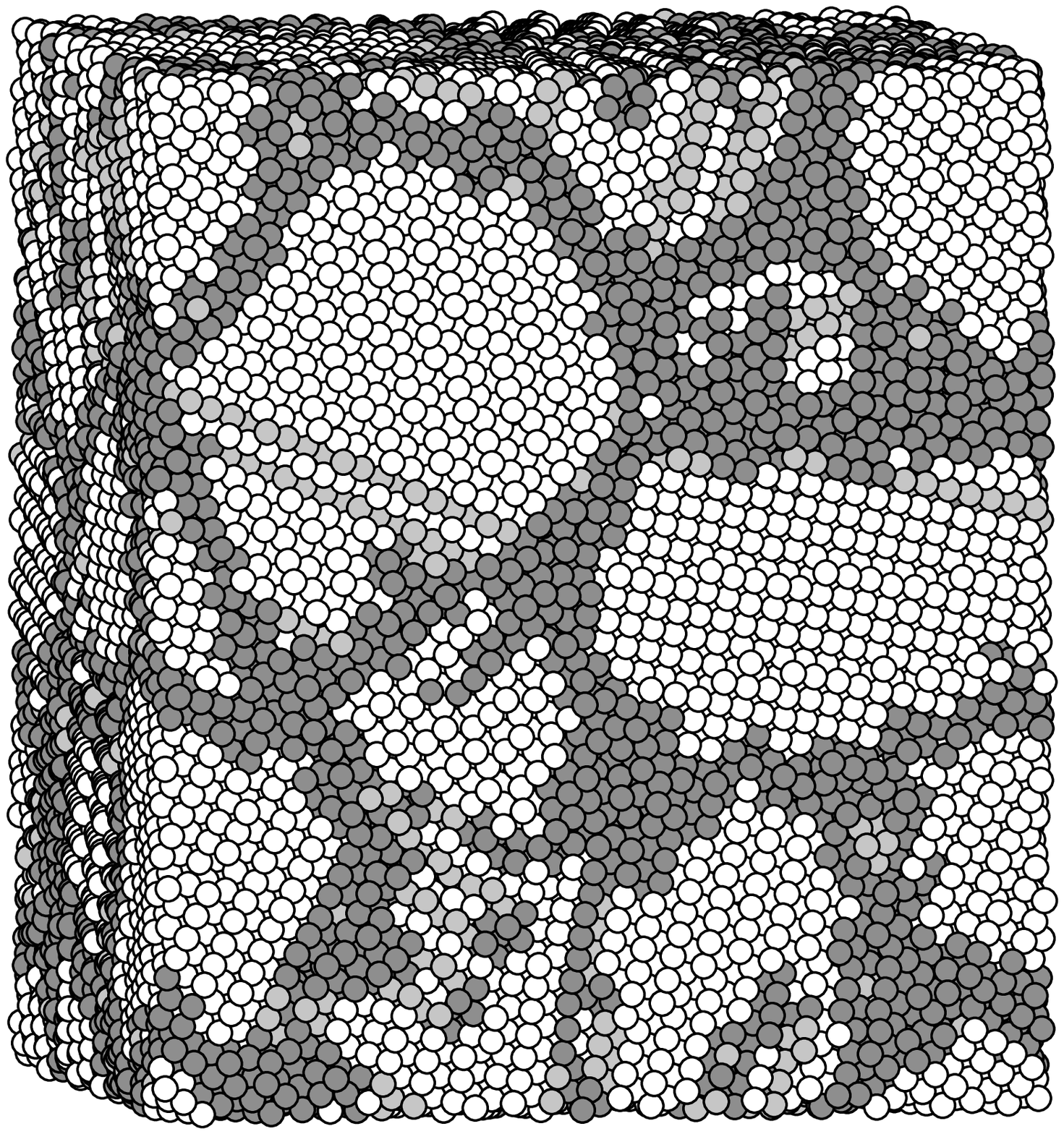, bb=42 56 511 534, clip=,
      width=0.49\linewidth}    
  \end{center}
  \caption{Simulation of the deformation of nanocrystalline copper at
    $T=300 K$ and $\dot\varepsilon = 5 \times 10^8 s^{-1}$. The four
    panels show the system after 0, 5, 7.5 and 10\% deformation.
    Little dislocation activity is seen.  In the second panel a
    Shockley partial dislocation (indicated by the arrow) is seen
    moving through the grain, leaving an intrinsic stacking fault
    behind.  Between 7.5 and 10\% deformation another Shockley partial
    has moved through the grain on an adjacent \{111\} plane,
    transforming the intrinsic stacking fault to an extrinsic one.}
  \label{fig:evolution}
\end{figure}

Figure \ref{fig:evolution} shows the deformation of a typical sample
with an average grain size of 5.21 nm.  The atoms have been color coded
according to the local crystal structure \cite{JoAn88,ClJo93}.  White
atoms are in local fcc order, and thus situated inside the grains.
Light grey atoms are in local hexagonal close-packed (hcp) order,
these atoms are at stacking faults.  Atoms in all other local
environments are colored dark grey.  These are typically atoms at grain
boundaries and in dislocation cores.

Some dislocation activity is seen in the system, as witnessed by the
generation of stacking faults.  The dislocation activity is
not sufficient to account for the observed plastic deformation.  A
detailed analysis of the deformation shows that the main deformation
mode is sliding in the grain boundaries \cite{ScDiJa98,ScVeDiJa98bx}.

\begin{figure}[tp]
  \begin{center}
    \epsfig{file=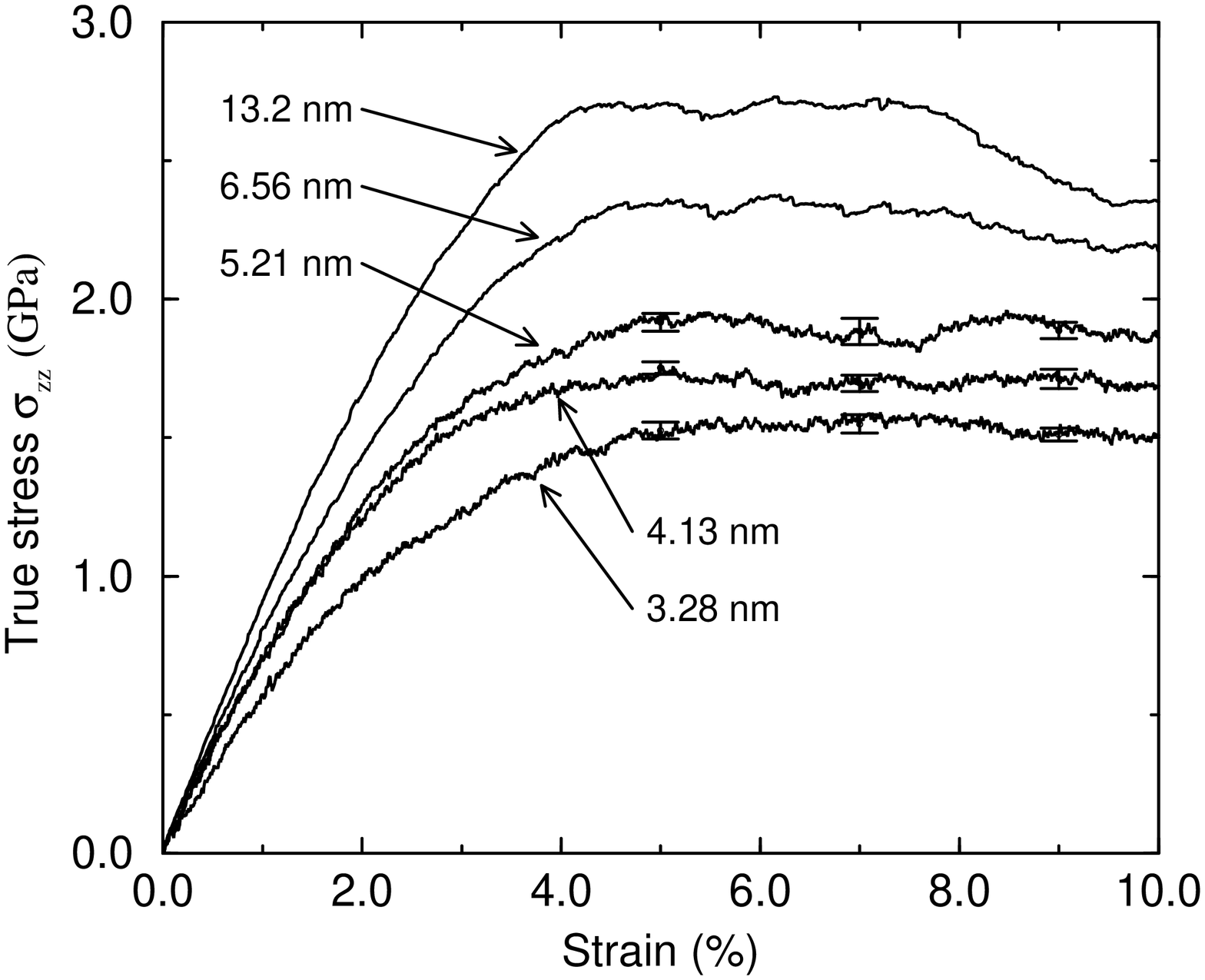, angle=-90, width=0.7\linewidth}
  \end{center}
  \caption{Stress-strain curves for nanocrystalline copper at 300K.
    Four simulations were performed at each grain size below 6\,nm,
    the curves shown are average stress-strain curves for the
    simulations.  The error bars are $1\sigma$ uncertainties on the
    mean value.  The larger grain sizes were simulated using a larger
    system (approximately $10^6$ atoms instead of $10^5$), only one
    simulation was made with $d = 6.56$\,nm, and two with
    $d=13.2$\,nm.  Adapted from Ref.~\protect\cite{ScVeDiJa98}.}
  \label{fig:stress}
\end{figure}

As the volume fraction of atoms in the grain boundary increases with
decreasing grain size, one would expect that increasing the grain size
increases the strength of the material as long as the grain boundaries
remain the carriers of the deformation.  This is indeed what we see in
the simulations.  Figure \ref{fig:stress} shows the stress-strain
curves obtained from simulations of systems with various average grain
sizes.  A ``reverse Hall-Petch effect'', i.e. a softening of the
material with decreasing grain size, is observed.

\subsection{Discussion}

The simulations of the deformation of nanocrystalline metals show a
reverse Hall-Petch effect in Cu and Pd for the grain sizes studied.
This softening of the material, when the grain size is reduced is
caused by plastic deformation in the grain boundaries.  There appears
to be two different deformation mechanisms active at different grain
sizes.  In metals with the very small grain sizes studied here, the
dominating deformation mechanism is sliding in the grain boundaries
through a large number of essentially uncorrelated events, where a few
atoms move with respect to each other at each event \cite{ScDiJa98}.

At much coarser grain sizes, dislocations are known to be the
dominating carriers of deformation.  In that regime, a hardening of
the material is seen, when the grain size is reduced, as the grain
boundaries act as barriers to the dislocation motion.  Experimentally,
this behavior is seen to continue far down into the nanocrystalline
range.

As the grain size is reduced, dislocation-mediated deformation becomes
more and more difficult.  On the other hand, the volume fraction of
the grain boundaries increases, favoring a deformation mechanism where
the grain boundaries carry the deformation.  Furthermore, as the grain
size is approaching the grain boundary thickness, it becomes
geometrically easier for slip to occur on more than one grain
boundary, without large stress concentrations where the grain
boundaries meet \cite{HaPa97}.

\begin{floatingfigure}{\narrowfigwidth}
  \noindent
  \epsfig{file=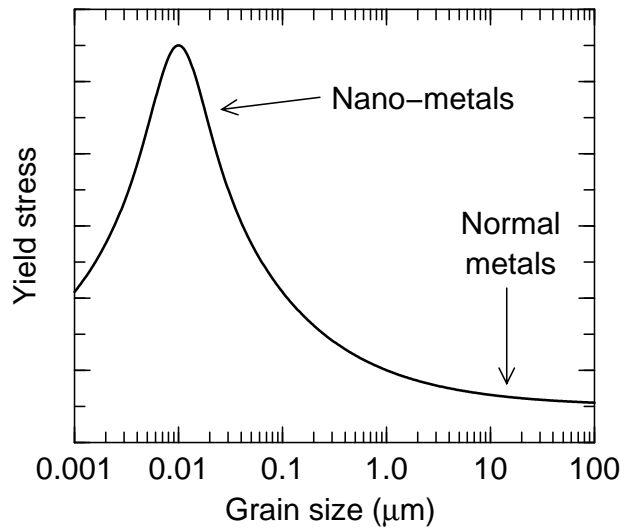, angle=-90, width=\narrowfigwidth}
  \caption{A \emph{qualitative} view of the relation between the yield
    stress and the grain size in metals.}
  \label{fig:maximum}
\end{floatingfigure}
The emerging picture is one where two deformation mechanisms compete.
One is active at very small grain sizes, another at larger grain sizes.
This leads to a maximum in the yield stress and hardness of
nanocrystalline metals at intermediate grain sizes, see Figure
\ref{fig:maximum}.

There have been many reports in the literature of a ``reverse
Hall-Petch effect'' at sufficiently small grain sizes.  However, the
hardness measurements are very sensitive to sample defects, and in
particular to sample porosity \cite{SaYoWe97,SaEaWe98,AgElYoHeWe98}.
In high-quality Cu samples, the Hall-Petch effect is seen to continue
at least down to grain sizes around 15\,nm \cite{SaYoWe97}.

There does not appear to be any \emph{unequivocal} experimental
evidence for a reverse Hall-Petch relationship in porosity free
nanocrystalline metals.  There are, however, indications of a
break-down of the ordinary Hall-Petch relation at grain sizes below
15\,nm in high quality copper samples produced by inert gas
condensation \cite{SaYoWe97}.  The ordinary Hall-Petch effect appears
to cease, although a reverse Hall-Petch relation is not seen.
It is still difficult to make direct comparisons between simulations
and experiment, as there is little overlap in grain size.  As the
experimental techniques improve, there is hope that more experimental
data will be gathered on high-quality samples with grain sizes below
10--15\,nm.  It should then become clear if the Hall-Petch effect does
indeed break down at these grain sizes.  At the same time, computer
simulations of larger systems might lead to observations of the
cross-over region between the reverse and the ordinary Hall-Petch
effect.  That cross-over region is beyond the reach of the largest
simulations presented here.  The grain boundaries remain the main
carriers of the deformation even when the grain size is increased to
13.2 nm, but be \emph{do} observe a slight increase in
dislocation activity as the grain size is increased.

\section{The effects of alloying}

As the main deformation mechanism is sliding in the grain boundaries,
it could be expected that altering the structure and composition of
the grain boundaries might have an effect on the properties of the
material.  One such modification is the addition of impurity
atoms in the grain boundaries.

\begin{floatingfigure}{\narrowfigwidth}
  \noindent
  \epsfig{file=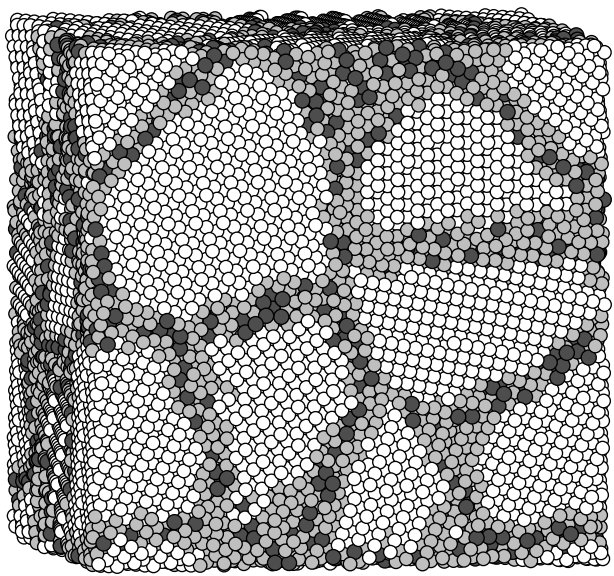, bb=47 62 505 489, clip=,
    width=\narrowfigwidth}
  \caption{Typical initial configuration for a silver-containing
    simulation.  White and light gray atoms are Cu atoms in the bulk
    and in grain boundaries, respectively.  Dark atoms are Ag atoms.}
  \label{fig:initAg}
\end{floatingfigure}
We have chosen to study the effects of silver impurities.  A reason
for choosing silver was that silver and copper are immiscible, low
concentrations of silver impurities in nanocrystalline copper can
therefore be expected to segregate to the grain boundaries.  We have
not studied the segregation process itself, as it is beyond the scope
of this study.  Segregation is a slow, diffusional process that cannot
be studied directly with MD simulations due to the timescales
involved.  Other approaches (such as Monte Carlo or Kinetic Monte
Carlo simulations) may be more appropriate.

Instead of simulating the segregation process, we have generated
systems that model nanocrystalline copper after such a segregation has
occurred.  It is done by replacing 25\% of the atoms in the grain
boundaries with silver atoms, see Figure \ref{fig:initAg}.  The system
is then annealed and deformed in the same way as the pure systems.

\begin{figure}[tp]
  \begin{center}
    \epsfig{file=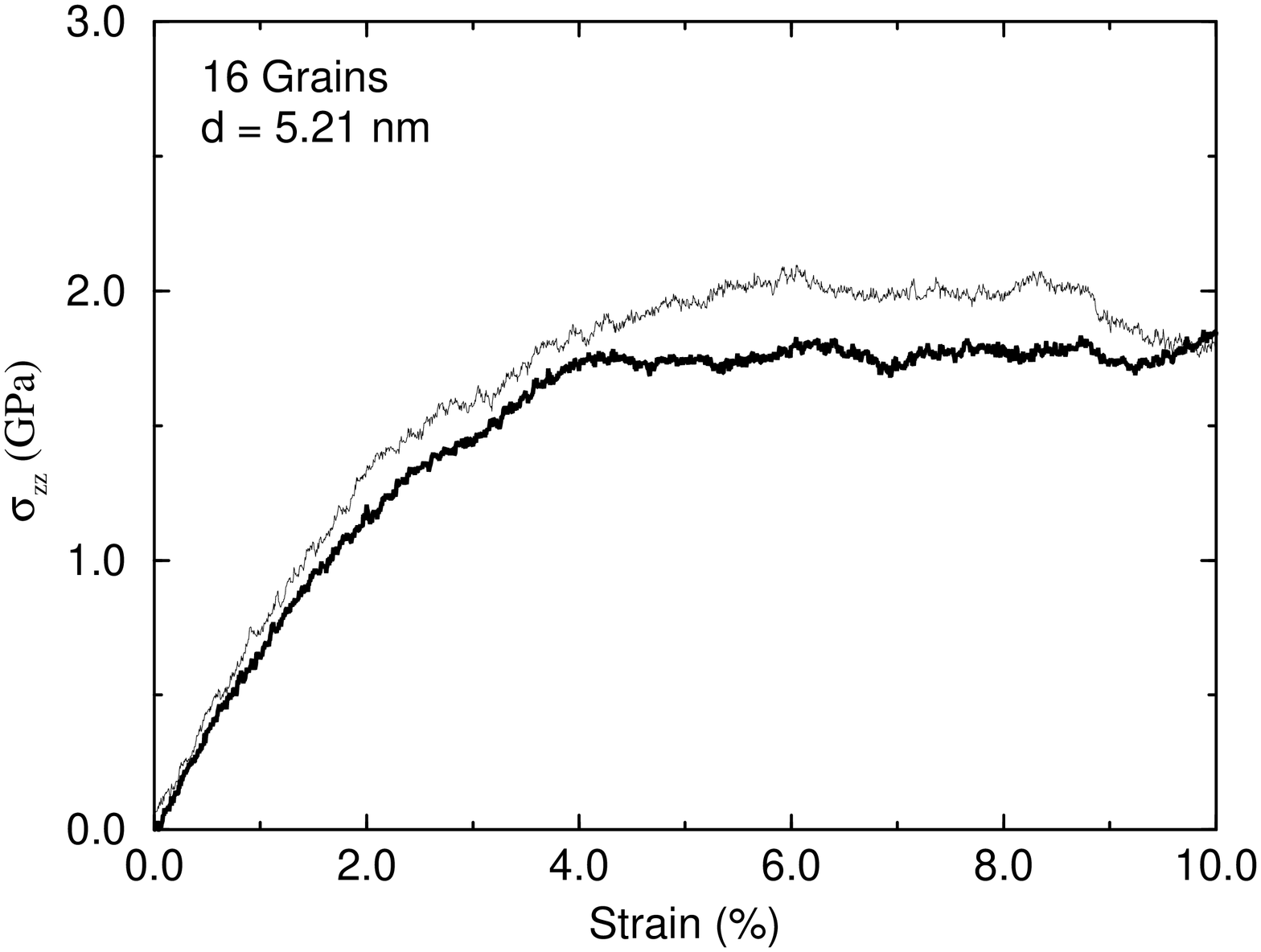, bb=96 56 567 680, angle=-90,
      width=0.45\linewidth}\hfill    
    \epsfig{file=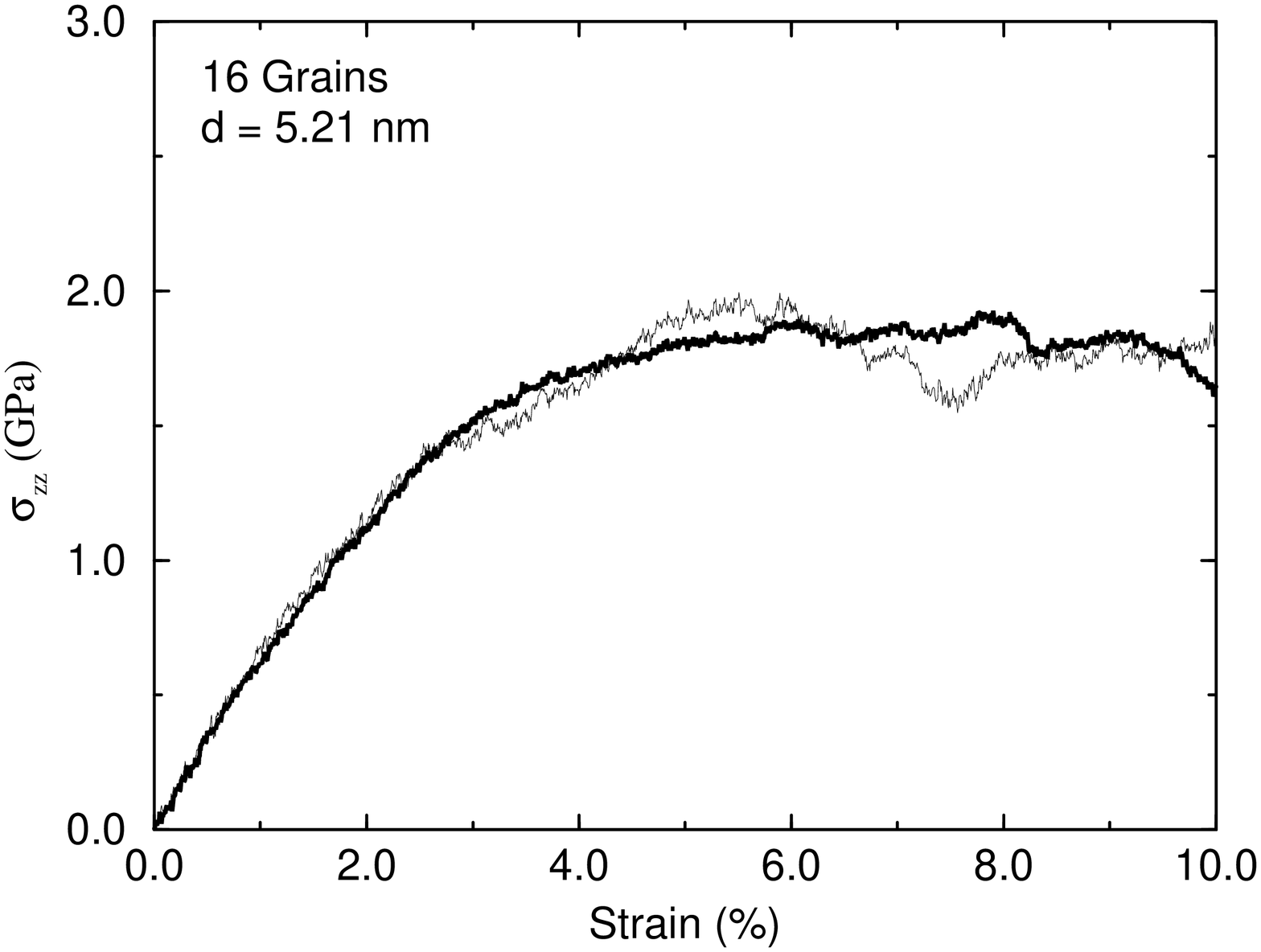, bb=96 56 567 680, angle=-90,
      width=0.45\linewidth}\\[\baselineskip]
    \epsfig{file=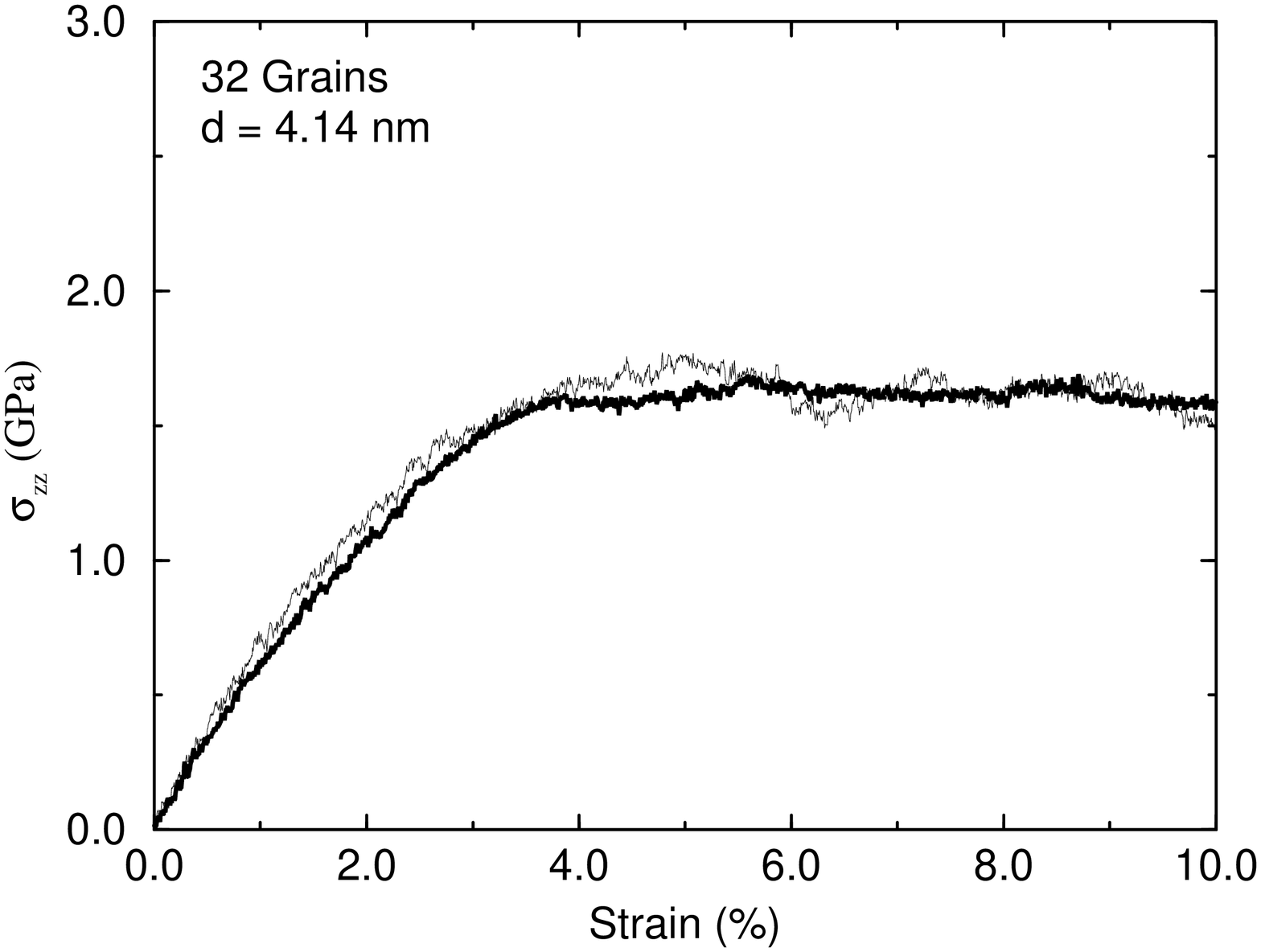, bb=96 56 567 680, angle=-90,
      width=0.45\linewidth}\hfill    
    \epsfig{file=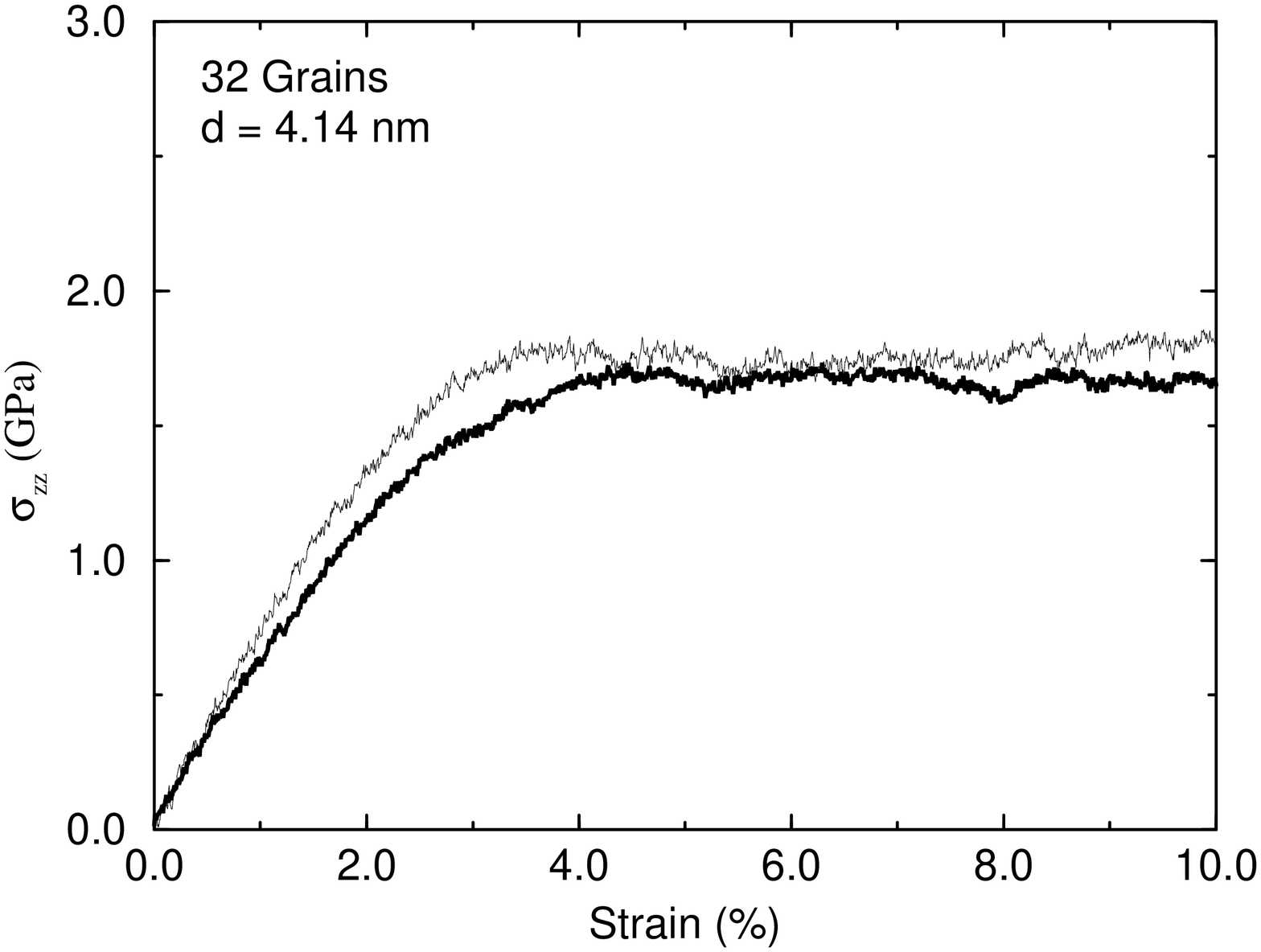, bb=96 56 567 680, angle=-90,
      width=0.45\linewidth}\\[\baselineskip]    
    \epsfig{file=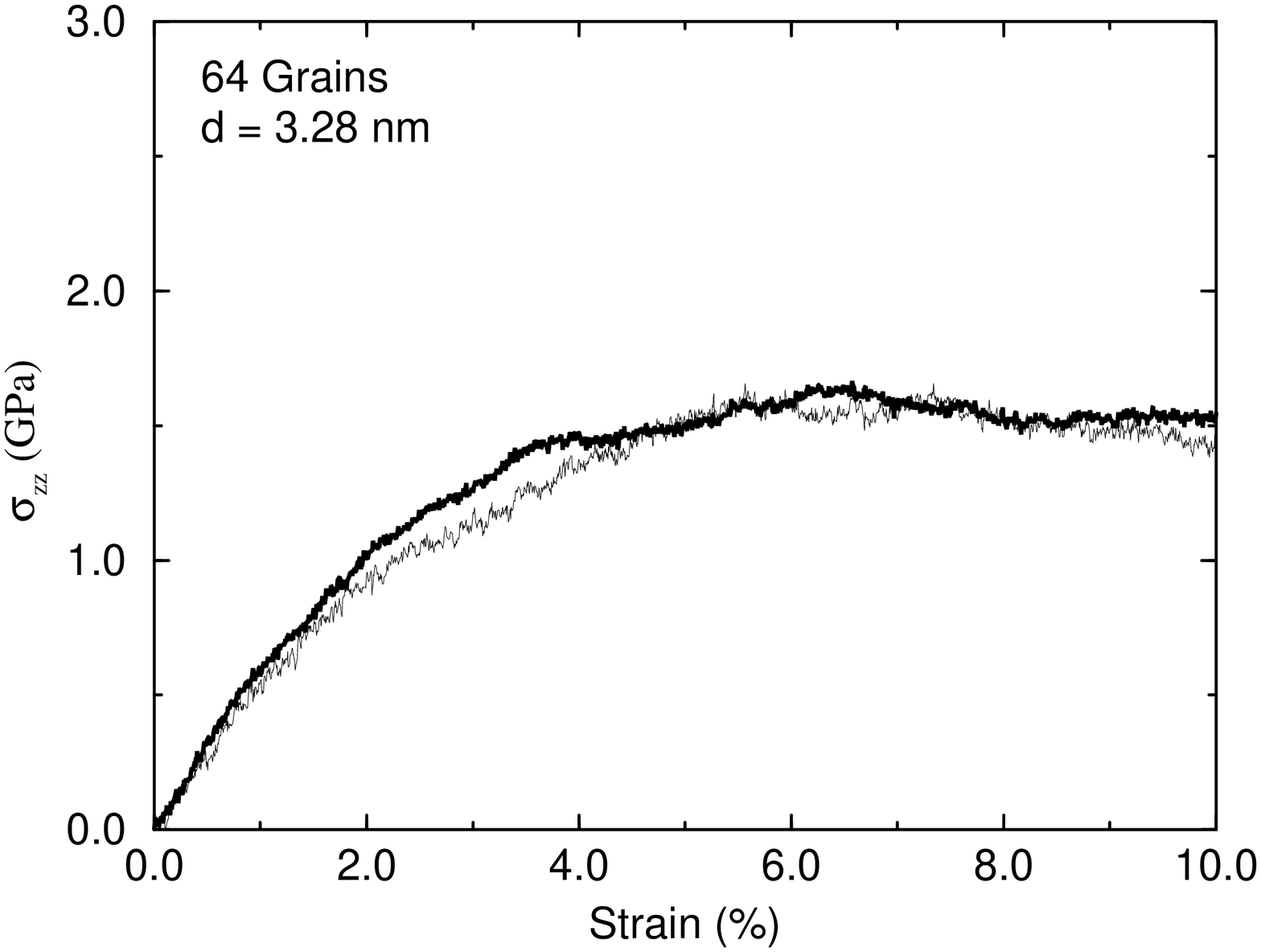, bb=96 56 567 680, angle=-90,
      width=0.45\linewidth}\hfill    
    \epsfig{file=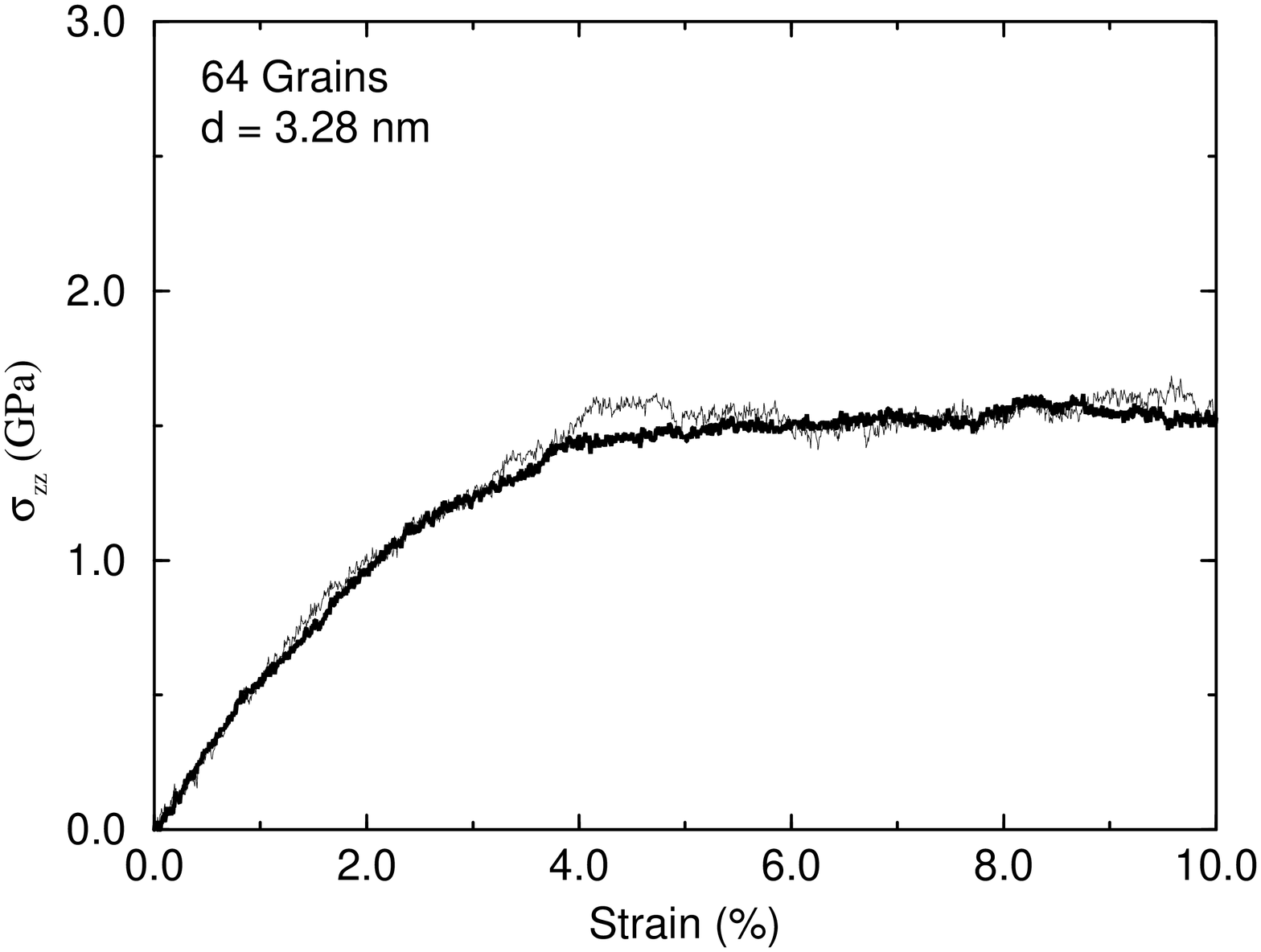, bb=96 56 567 680, angle=-90,
      width=0.45\linewidth}    
  \end{center}
  \caption{Comparison between simulations with and without silver
    atoms in the grain boundaries.  Each graph shows the stress-strain
    curve of a simulation with 25 at-\% Ag in the grain boundaries
    (thick lines), and the same system (i.e.\ the same grain
    structure) without the silver (thin lines).  For each grain size,
    two different sets of simulations were performed with different
    random grain structures.}
  \label{fig:compare}
\end{figure}

\begin{figure}[tp]
  \begin{center}
    \epsfig{file=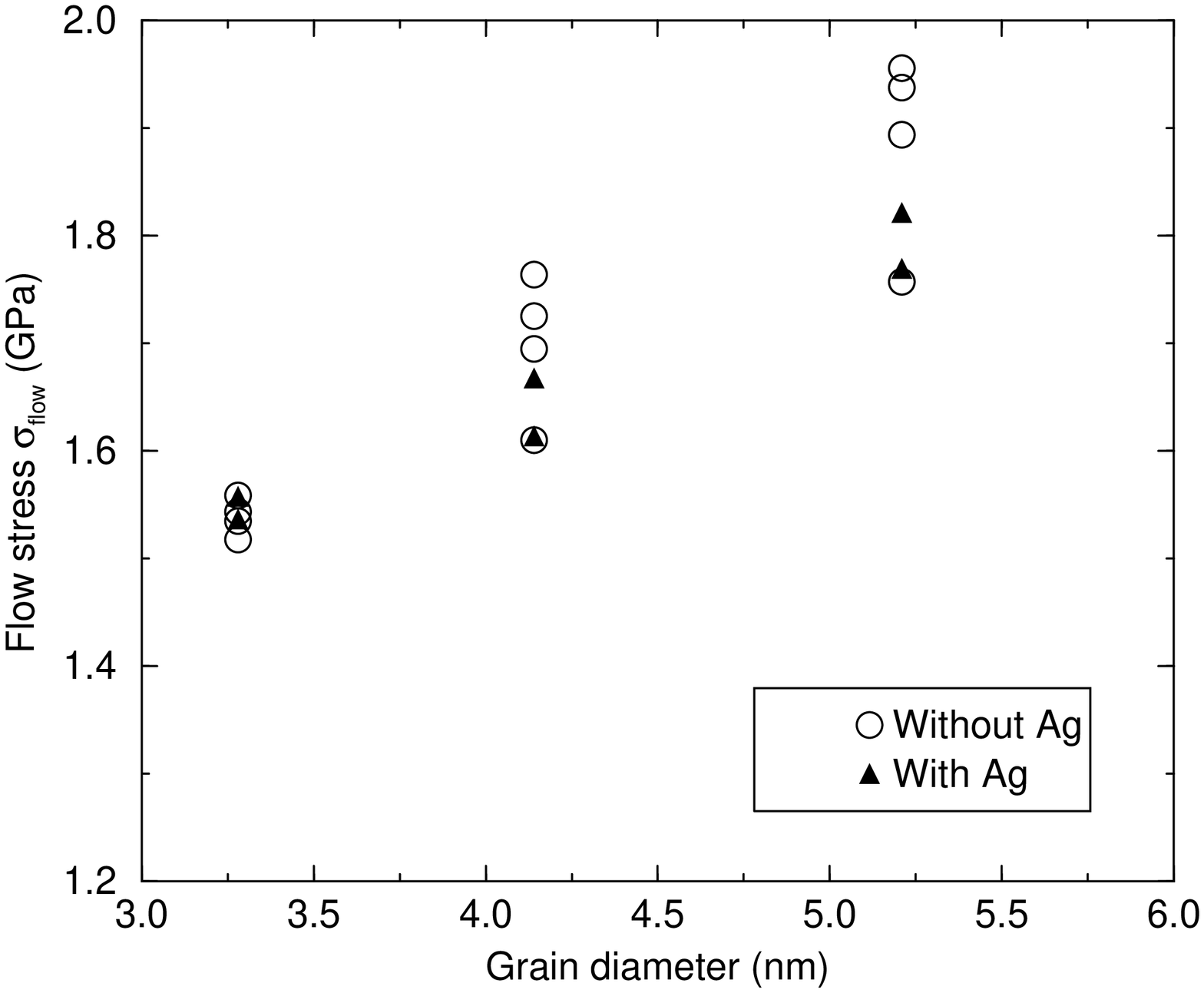, angle=-90, width=0.7\linewidth}    
  \end{center}
  \caption{The flow stress for the simulations with (triangles) and
    without (circles) silver in the grain boundaries, as a function
    of the grain size.  Each point represents one simulation.}
  \label{fig:flowstress}
\end{figure}

Figure \ref{fig:compare} shows how the stress-strain curves have
changed, when silver is introduced in the grain boundaries.  The
general trend seems to be a slight softening of the material, although
the effect is very weak and not seen in all systems.  Figure
\ref{fig:flowstress} shows the flow stress levels for the different
simulations, i.e.\ the stress level at the horizontal part of the
stress-strain curve (for simplicity the flow stress was defined as the
average stress for $6\% \le \varepsilon \le 10\%$).  We again see the
tendency for the silver-containing systems to be softer.

In the simulations of impurities in the grain boundaries, the
mechanical behavior is close to what is seen in the pure systems.

As the major part of the deformation happens in the grain boundaries,
one could expect that adding impurities in the grain boundaries could
have a relatively large effect on the mechanical properties.  The
simulations presented here show that this is not the case, when silver
is used as impurities in copper.  We see a tendency towards a
weakening of the material, but the effect is barely detectable.

The atomic bonding in copper and silver are of a similar nature, and
the size of the atoms are not very different.  This may account for
the absence of a stronger effect of alloying in the grain boundaries.

\section{Conclusion}

Atomic-scale simulations have been used to study the deformation
mechanisms in nanocrystalline copper with and without impurities in
the grain boundaries.  In both cases, we find that the main
deformation mode is sliding in the grain boundaries through a large
number of apparently uncorrelated events, each involving only a few
atoms.  Some dislocation activity is seen in the grains, the
dislocations are probably necessary to allow the grains to deform a
little, as they glide past each other.

We observe a \emph{reverse Hall-Petch effect}, i.e.\ a hardening of
the material as the grain size is increased: as the amount of grain
boundary atoms is decreased, the deformation becomes harder.  At the
same time the dislocation activity is seen to increase a little with
grain size.  This effect is seen in the entire range of grain sizes
that we have studied (3 to 13 nm), but at some point we expect that
dislocation motion will begin to dominate the behavior of the system.
When that happens, the yield strength should begin to \emph{decrease}
with increasing grain size.

Adding silver to the grain boundaries has only a weak effect on the
properties of the material.  The strength of the material is seen to
decrease marginally, when the silver is introduced.  We expect that
other elements, which chemically behave in a way that is more different
from copper will have a larger effect, but such simulations remain to
be made.

\section{Acknowledgments}

This work was financed by The Danish Technical Research Council
through Grant No.\ 9601119.  Parallel computer time was financed by the
Danish Research Councils through Grant No.\ 9501775.  Center for
Atomic-scale Materials Physics is sponsored by the Danish National
Research Council.

\renewcommand{\refname}{REFERENCES}

\end{document}